\documentclass[a4paper,fleqn,usenatbib]{mnras}
\usepackage{graphicx}
\usepackage{natbib}
\usepackage{aas_macros}
\usepackage{amssymb}
\usepackage{ulem}
\usepackage{etoolbox}
\makeatletter

\patchcmd{\NAT@citex}
  {\@citea\NAT@hyper@{%
     \NAT@nmfmt{\NAT@nm}%
     \hyper@natlinkbreak{\NAT@aysep\NAT@spacechar}{\@citeb\@extra@b@citeb}%
     \NAT@date}}
  {\@citea\NAT@nmfmt{\NAT@nm}%
   \NAT@aysep\NAT@spacechar\NAT@hyper@{\NAT@date}}{}{}

\patchcmd{\NAT@citex}
  {\@citea\NAT@hyper@{%
     \NAT@nmfmt{\NAT@nm}%
     \hyper@natlinkbreak{\NAT@spacechar\NAT@@open\if*#1*\else#1\NAT@spacechar\fi}%
       {\@citeb\@extra@b@citeb}%
     \NAT@date}}
  {\@citea\NAT@nmfmt{\NAT@nm}%
   \NAT@spacechar\NAT@@open\if*#1*\else#1\NAT@spacechar\fi\NAT@hyper@{\NAT@date}}
  {}{}
\def\comment{\colorbox{yellow}}{}
 \def\ASK\comment{ASK}

\makeatother

\title[Decompositions of low-mass galaxies]{How similar is the stellar structure of low-mass late-type galaxies to that of early-type dwarfs?}

\author[Janz et al.]{J.~Janz$^{1}$\thanks{Email: jjanz@swin.edu.au},  E.~Laurikainen$^{2}$,  J.~Laine$^2$, H.~Salo$^2$, \& T.~Lisker$^3$\\ $^1$Centre for Astrophysics and Supercomputing, Swinburne University, Hawthorn, VIC 3122, Australia\\
$^2$Astronomy Research Unit, PO Box 3000, University of Oulu, FI-90014 Oulu, Finland\\
$^3$Astronomisches Rechen-Institut, Zentrum f\"ur Astronomie der Universit\"at Heidelberg, M\"onchhofstra\ss e 12-14, D-69120 Heidelberg, 
Germany}

\begin{document}

\pagerange{\pageref{firstpage}--\pageref{lastpage}} \pubyear{2015}

\maketitle

\begin{abstract}

We analyse structural decompositions of 500 late-type galaxies (Hubble $T$-type $\ge 6$) from the  \textit{Spitzer Survey of Stellar Structure in Galaxies (S$^4$G; Salo et al.)}, spanning a stellar mass range of about $10^7$ to a few times $10^{10}$ M$_\odot$. Their decomposition parameters are compared with those of the early-type dwarfs in the Virgo cluster from Janz et al. They have morphological similarities, including the fact that the fraction of simple one-component galaxies in both samples increases towards lower galaxy masses. We find that in the late-type two-component galaxies both the inner and outer structures are by a factor of two larger than those in the early-type dwarfs, for the same stellar mass of the component. While dividing the late-type galaxies to low and high density environmental bins, it is noticeable that both the inner and outer components of late types in the high local galaxy density bin are smaller, and lie closer in size to those of the early-type dwarfs. This suggests that, although structural differences between the late and early-type dwarfs are observed, environmental processes can plausibly transform their sizes sufficiently, thus linking them evolutionarily.

\end{abstract}


\label{firstpage}
\section{Introduction}
At low galaxy masses the analogue to Dressler's (\citeyear{dressler})
morphology density relation \citep{Binggeli87} in clusters led to the inference
that late-type galaxies are transformed to early-type dwarfs by
environmental processes \citep[e.g.][]{Boselli06}.  In recent years,
the suggested processes, i.e. starvation, ram pressure stripping,
and harassment (\citealt{Larson80,LinFaber,Moore}), and their relative
importance have been actively discussed. Studies focused on various aspects:
the relative contribution to the total population of early-type dwarfs
systems (e.g.~\citealt{Boselli}), the time of the transformation, i.e.~late infall versus
formation on cosmological time-scales  along with cluster assembly (e.g.~\citealt{Boselli,lisker09}), 
and its location, i.e.~pre-processing in
groups versus formation in galaxy clusters (e.g.~\citealt{lisker13,Paudel}).

The early and late-type dwarfs differ in two essential aspects: Their star formation activity 
and optical morphology.  Quiescent early types are basically absent in low density
environments below a mass threshold
\citep{2012ApJ...757...85G}. Within galaxy clusters star-forming
late-type dwarfs and early-type dwarfs with recent residual star
formation are preferentially found in the outskirts of the cluster
\citep{Binggeli87,lisker07}.  The same applies to disc features  and disc-like kinematics in
early-type dwarfs \citep{lisker07,toloba}.  These trends were used to underpin the 
idea of a environmentally driven transformation from late to early type.

\citet{2012ApJ...745L..24J} applied the two-dimensional decomposition
method to early-type dwarf galaxies in the Virgo cluster. Both single
component S\'ersic fits and more complicated multicomponent
decompositions were performed, and it turned out that many of the
galaxies are better modelled with the multicomponent models.
\citet{2013arXiv1308.6496J} concluded that the decompositions of those
galaxies are likely to represent discs with a mass distribution more
complex than a single component, but not as distinct as a separation
into a bulge and a disc with clearly different physical
properties. However, they were not able to do any thorough comparison
to late-type galaxies \textit{of the same mass}, since late-type
spiral and irregular galaxies (Hubble types 6 $\le T \le$ 10, i.e.~Scd
and later) are under-represented in typical samples of decomposition
studies in the Local Universe
(e.g.\ \citealt{2008MNRAS.tmp..752G,2010MNRAS.405.1089L}, but see,
e.g., \citealt{Boker:2003ju,barazza+08}). 

Such a
structural comparison is essential for addressing the question whether
late-type galaxies as we observe them today can be the progenitors of
early-type dwarfs, and, if this is the case, for assessing how much
their structure needs to be altered during the transformation.
Recent progress in the structural decomposition of
low-mass galaxies
\citep{2012ApJ...745L..24J,2013arXiv1308.6496J,arXiv:1503.06550}
enables us to quantitatively compare the structures underlying the
morphological differences between early and late-type dwarfs.

In this Letter we analyse the decompositions of
near-infrared light distributions of 500  late-type
galaxies   \citep{arXiv:1503.06550} from the {\textit Spitzer} Survey of Stellar Structure in Galaxies
\citep[S$^4$G;][]{2010PASP..122.1397S}, 
and compare them to a similar data set for early-type dwarfs \citep{2013arXiv1308.6496J}.
We find both similarities and
differences between these two galaxy types, and discuss  processes
that may bridge them.

\section{Sample and Data}
The S$^4$G  is a magnitude
($B_{T,\textrm{corr}}<15.5$ mag), size ($D_{25}>1\arcmin$), and volume
($d<40$ Mpc, using distances based on H~\textsc{i}  measurements) limited Spitzer
survey of over 2300 galaxies. S$^4$G provides images at 3.6 and 4.5
$\mu$m, having a depth equivalent to about $1 M_{\sun}$ pc$^{-2}$ at
the average distance of the galaxy sample.\footnote{This corresponds
  to $\mu_H\ge$ 25 mag arcsec$^{-2}$ for typical stellar populations
  of the early-type dwarfs in the comparison sample, and is deeper
  than the typical Virgo dwarf images,  which had total integration times
  so that comparable signal to noise at the half-light radius was achieved.}
  This makes the S$^4$G ideal for the challenges
posed by galaxies at the end of the Hubble sequence, due to the
low galaxy masses and typically also irregular star-forming
regions.   The effective image resolution is $2\farcs1$ full width at half-maximum (see \citealt{arXiv:1503.06550}). 
The sample is particularly representative 
for the gas rich late-type galaxies at the faint end of the luminosity
function considered here.

 We use the morphological classifications from \citet{Buta} to select
 late-type galaxies with a Hubble $T$-type $\ge 6$. The structural
 decompositions for these galaxies are taken from
 \citet{arXiv:1503.06550}. Only decompositions with an inclination
 $i<65^\circ$ and flagged as high quality 
 are further considered.  
 The final sample contains $\sim$500 galaxies.  The
 S$^4$G decompositions employed \textsc{galfit}
 \citep{2010AJ....139.2097P}, relying on parametric fits and the
 statistical uncertainties of the pixels, and were evaluated with
 \textsc{galfidl} visualisations.\footnote{H.~Salo,
   \url{http://cc.oulu.fi/~hsalo/galfidl.html}.}  Galaxies with
 multiple components were fitted using an exponential function for the
 outer structures (in few cases a S\'ersic function), and Ferrers or
 S\'ersic functions for the inner components (+PSF for a possible
 central point source).  For the comparison with the early-type
 dwarfs, we convert the 3.6 $\mu$m magnitudes in S$^4$G to the
 $H$-band, using an average colour calculated from S$^4$G and the 2MASS
 extended source catalogue \citep[XSC;][]{Jarrett:2000fz}.  The converted $H$-band
 magnitude has a tight relation with the stellar masses based on the 3.6 $\mu$m magnitudes of
 \citeauthor{mm2015} (\citeyear{mm2015}; $\log M_* \approx -0.39\ M_H + 1.61$).
   The
 comparison sample of early-type dwarf galaxies in the Virgo cluster
 \citep[galaxies with early-type classification in the Virgo cluster
   catalogue, \citealt{Binggeli}, with $M_r>-19$
   mag]{2013arXiv1308.6496J}  used a similar \textsc{galfit} approach and
the same limit in galaxy inclination.

\section{Analysis}
\label{section:analysis}
\begin{figure}  
   \centering
   \includegraphics[height=0.44\textwidth,angle=-90]{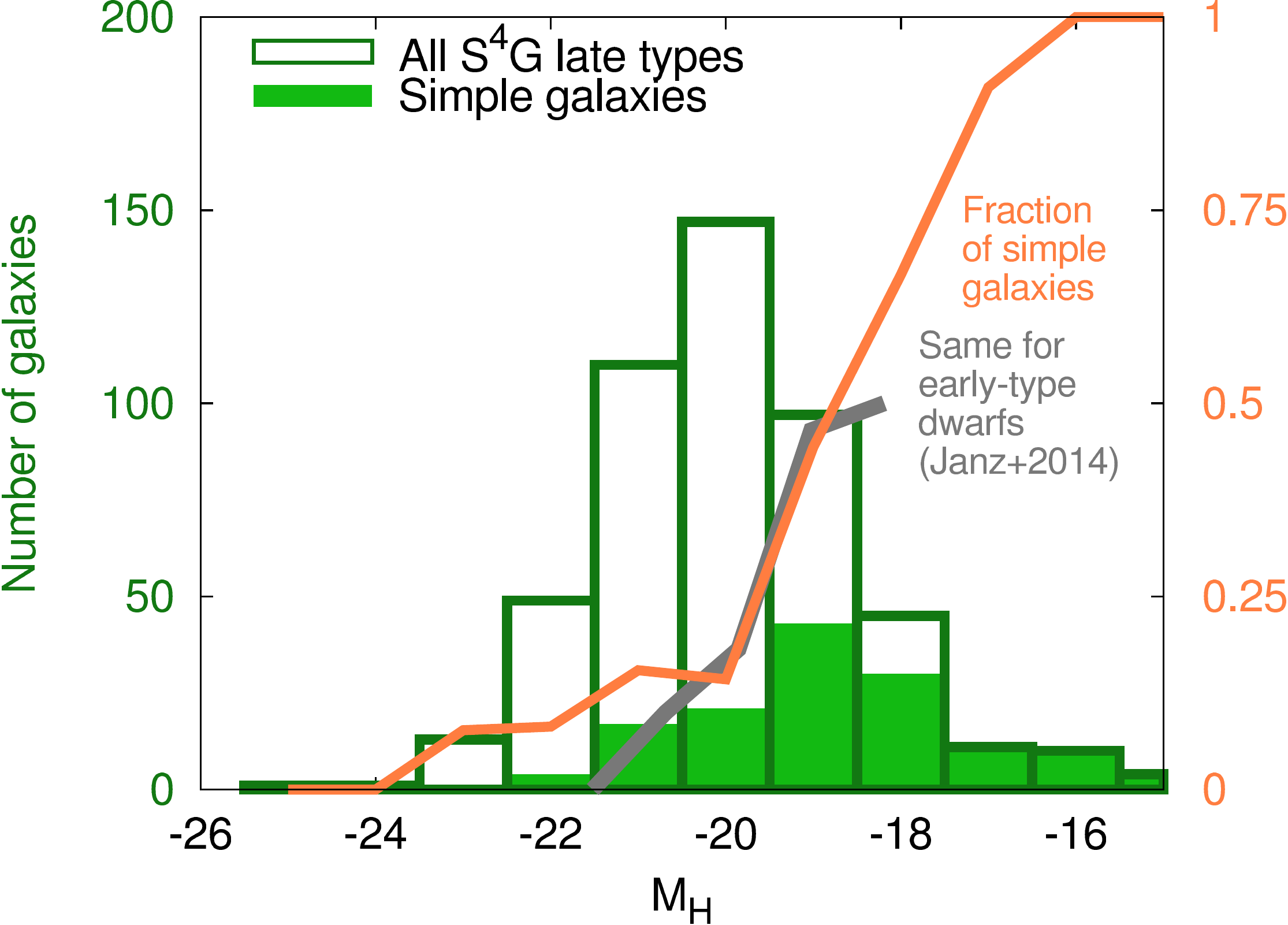} 
      \caption{Histogram of brightnesses of analysed galaxies. Open
        bars represent all galaxies (with S$^4$G decomposition and an
        ellipticity $< 1-\cos 65^\circ$), while the solid bars
        indicate simple galaxies (see text).  The orange line shows
        the fraction of simple galaxies, with numerical values
        indicated in the right $y$-axis. For comparison the
        corresponding fraction for the early-type dwarf galaxies in
        the Virgo cluster \citep{2013arXiv1308.6496J} is displayed
        as a grey line.}
   \label{fig:fraction}
\end{figure}

We divide the  decompositions of late-type galaxies into simple and
multicomponent systems, in a similar manner as \citet{2013arXiv1308.6496J} did for the early-type dwarfs: Simple galaxies are those 
for which a good fit is achieved with a single  component, and,
if necessary, a central point source.\footnote{ For the late-type galaxies in the S$^4$G sample
the exponential function was used for the simple galaxies, while \citet{2013arXiv1308.6496J} 
used a more flexible S\'ersic function. The definitions of simple galaxies are still comparable, since
the simple early-type dwarfs are close to exponential as \citet{2013arXiv1308.6496J} state.
This is also confirmed by both simple galaxies showing compatible S\'ersic $n$ vs $M_H$ distributions when using 
the one component S\'ersic fit for the late types, which was also carried out for all galaxies in the S$^4$G.} 
By multicomponent
galaxies we mean systems in which two (fairly shallow) components,
presumably forming part of the disc, were fitted with separate
functions.

We find the fraction of simple late-type galaxies in S$^4$G
to strongly depend on galaxy luminosity so that it
increases towards fainter galaxies (Fig.~\ref{fig:fraction}).  This luminosity dependence is 
remarkably similar to that previously obtained for the
early-type dwarfs in the Virgo cluster \citep{2013arXiv1308.6496J}.

Next the parameters of the multiple component galaxies are compared in the
two samples  (Fig.~\ref{fig:nirs0s_env}).  Unlike bulges and discs in
bright spiral and lenticular galaxies \citep[from bulge-disc-bar structural
 decompositions;][]{2010MNRAS.405.1089L}, the components in
late-type and early-type dwarfs do not follow any tight scaling
relations. They both appear to scatter more randomly.  However, there is a
clear difference in size between their structure components:
For a given mass of a structure component, those in the late-type
systems are on average twice as large as those in the early-type dwarfs.
The overlap between the early and late-type
dwarfs in Fig. 2 is restricted to a relatively small number of
objects.

In order to investigate whether the environmental effects could
  play a role in the above difference between the early and
  late-type dwarfs, we study the local environments of S$^4$G galaxies
 as quantified by \citet{laine}: the
XSC and the 2MASS Redshift Survey \citep[RSC;][]{huchra} are used to
calculate the projected surface number density $\Sigma_3 =
{{3}\over{\pi R_3^2}},$ with the projected distance $R_3$ to the third
nearest neighbour galaxy in a velocity interval of $\pm 1000$ km s$^{-1}$
around the primary galaxy.\footnote{Due to the RSC completeness
  limits, the availability of the environment information steeply
  decreases towards fainter magnitudes, with $\sim$50\% of the
  galaxies covered at $M_H\sim-19.5~\textrm{mag}$.}  We divide the
sample galaxies to low and high-density environments using the median
value of the local density of all S$^4$G galaxies with this
information available ($\log \Sigma_3=0.47$). Interestingly, the sizes of the outer
components are smaller for those late-type dwarfs located in high local galaxy
density environments: as much as about half of the galaxies in the
high-density sub-sample have parameters similar to those of the
early-type dwarfs, located in the high-density environment of the
Virgo cluster.

\section{Discussion}
\subsection{Frequency of structures as function of luminosity}
Here we found that the fraction of the
structurally simple late-type galaxies in S$^4$G increases with 
decreasing galaxy luminosity, and hence stellar mass, in a similar
manner as among the early-type dwarfs in the Virgo cluster. This rise
of the fraction was previously found also with profile decompositions
by \citet{2001A&A...372...29G}. 
This trend is also paralleled by the frequency of
bars, spirals, and inner discs in early-type
dwarfs, detected  by \citet{liskera} using unsharp masking, which
also become scarcer at lower mass.

For stellar discs the ability to form and retain substructures depends
on the Toomre stability parameter \citep{1964ApJ...139.1217T}. It
is proportional to the random motions within the disc, for a given
rotation curve, 
and inversely proportional to the
surface mass density.  The sharp decline of the fraction of bars,
spiral arms, and other multicomponent structures towards lower galaxy
mass {inside the dwarf  regime} may therefore be explained by the low-mass discs being more
  stable.
  In addition to the lower surface density in low-mass discs, there is
  also observational evidence for them being dynamically hotter
  \citep{2005AJ....130.1574S,Yoachim,2010MNRAS.406L..65S}.
  Theoretically, this may be expected from processes during their
  formation and internal evolution
  \citep[e.g.][]{2007MNRAS.382.1187K}.  Likewise external processes,
  such as tidal interactions and interactions with hot gas
  \citep{2003ApJ...589..752G,Smith:ey,Smith:2012il}, can form such a
  trend by heating the low-mass galaxies with  shallow potential
  wells more efficiently.  Whether this qualitative explanation holds,
  will need further testing, since additional factors  (see also \citealt{rsj2016}) such as
  systematic differences in the stellar mass-to-light ratio, as well as
  differences in the expected dark matter content,   from late to early-type dwarfs play a role.

\subsection{Galaxies with multiple components}

\begin{figure*}
   \centering
   \includegraphics[height=0.91\textwidth,angle=-90]{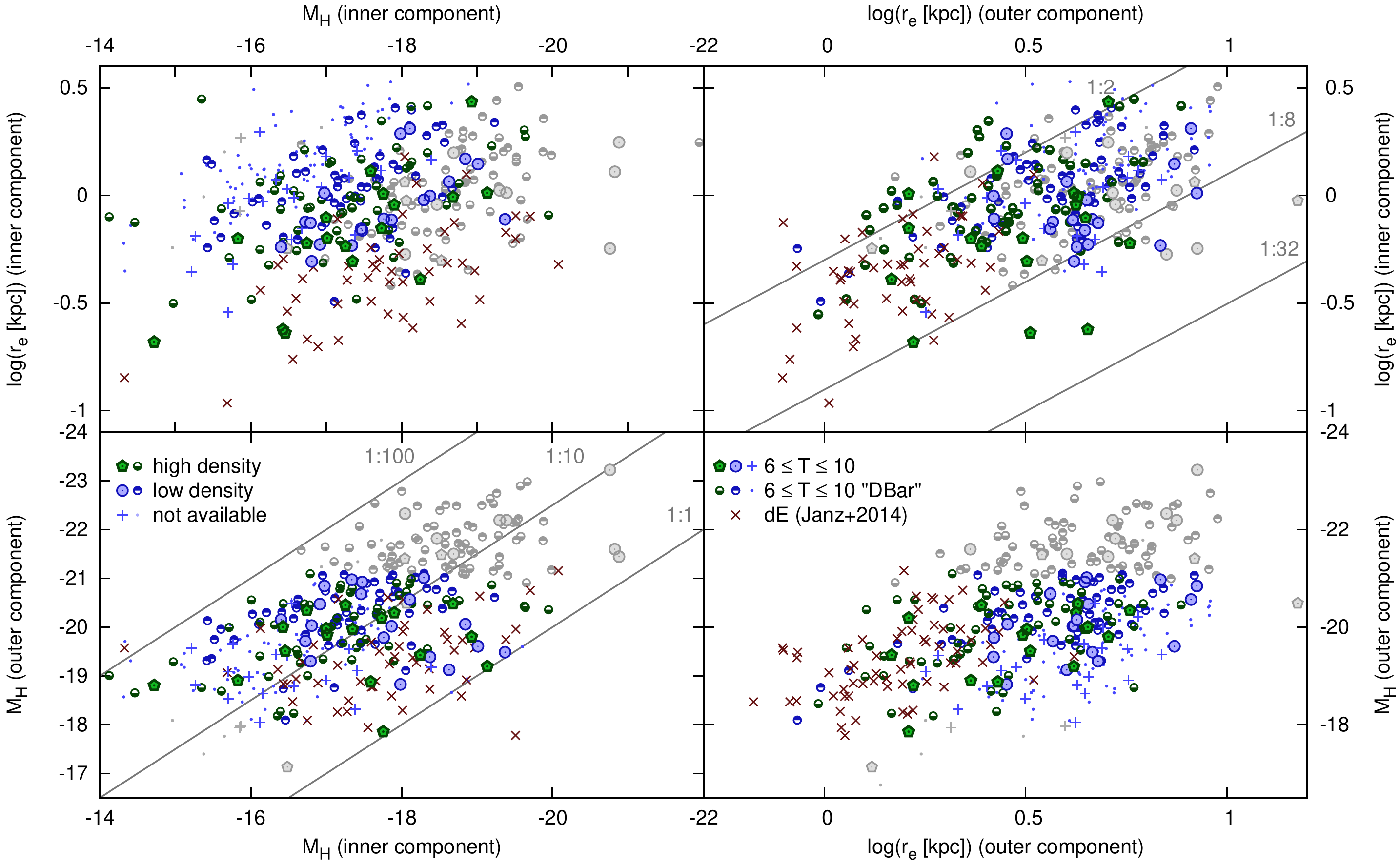} 
   \caption{Inner and outer components  of the multicomponent galaxies are shown in diagrams of the various combinations of their sizes and brightnesses.  
  The late-type galaxies are subdivided  into low ({\it blue filled symbols}) and high ({\it green colour}) local galaxy density environments (galaxy with lacking environment information are plotted with {\it blue crosses} and {\it points}). A further distinction is made whether or not the inner component is called bar in the S$^4$G decomposition (see text and the warning about this classification there). 
In addition we include galaxy decompositions for  early-type dwarf galaxies  \citep[\textit{red crosses,}][]{2013arXiv1308.6496J}.
 The late-type galaxies with a total $H$-band magnitude outside the range spanned  by the early-type dwarfs are plotted with grey symbols.
 Differences between the two populations and trends with environment for the late-type galaxies are described in Section \ref{section:analysis}.
  }
   \label{fig:nirs0s_env}
\end{figure*}

A general similarity of the structural parameters between late  and early-type dwarf galaxies was recently highlighted by, e.g., \citet{Meyer:2014bw}, \citet{Young14}, \citet{Mahajan15}, and \citet{lian15}. However, here we find 
differences of the decomposition parameters that seem large enough to argue that a simple fading of late-type galaxies cannot account for the structures  in early-type dwarfs (Fig.\ \ref{fig:nirs0s_env}). Despite the difference in size, the decompositions of both galaxy types share the characteristic of a lack of clear scaling relations as found for those in lenticulars and early-type spirals, and their parameters are offset from the extrapolation of those relations (e.g.~when plotting other galaxy types in S$^4$G or when comparing to \citealt{2010MNRAS.405.1089L}).  The scales of bulges and discs in these systems  are generally also better separated, i.e.~many having bulge-to-disc scale ratios between 1:8 and 1:32, and only massive early types have bulges with scales comparable to those of their discs.

A large majority of  multicomponent late-type dwarfs are systems that 
 have elongated inner features, in $>$50\% of the galaxies a bar
 component was fitted.
  While \citet{arXiv:1503.06550} warn not to rely on the naming convention
of their decomposition types when identifying bars, this is a noteworthy difference when comparing
to the early-type dwarfs with  bar components fitted in only 14\% of the galaxies.
\citet{arXiv:1503.06550}  also point out that a low contrast between the
inner and outer components in some of the galaxies did not allow
fitting the bar with a separate function, even if such a bar was
recognized in the visual classification by \citet{Buta}, who identified bars in   80$\%$ of  galaxies at inclinations of  $i<$
60$^{\circ}$. This is manifested also in
a larger bar fraction in the visual classification, compared to that
obtained from ellipse fitting, or while using Fourier amplitudes 
of density to identify bars (see \citealt{diaz-garcia}). 
Those late-type galaxies with multiple components, 
comprising a disc and inner elongated structure (`DBAR' in \citealt{arXiv:1503.06550}), are included
in Fig.~\ref{fig:nirs0s_env}, but with different symbols. They do not change the conclusion of a shift 
between the parameter spaces occupied by late and early-type galaxies.

The bars in  low-mass late types are clearly unlike bars in
bright galaxies. Therefore, also the physics of the bar
formation and possible destruction are expected to be different. 
Bars in late-type systems are not centrally concentrated  \citep{arXiv:1503.06550}, and
lack vertically thick inner regions typically found in
 Milky Way mass galaxies ($\sim$50\% of bars), which are manifested as
Boxy/Peanut/X-shaped structures and barlenses (\citealt{Athanassoula,2014MNRASL}; see also
\citealt{Combes}).

\subsection{Are early-type dwarfs transformed late types?}
The similarity of behaviour as a function of luminosity of the fractions of 
 structurally simple galaxies
 for early and late-type galaxies  could be interpreted as  early-type dwarfs being compatible
with being transformed late-type systems. However,  we argued in Section 4.1 that
this may rather be driven by changes of disc stability. 
Furthermore, the twice as large sizes of  inner and outer components in late-type dwarfs,
when compared to those of the early types, seems to argue against
such a connection. Likewise, the frequency of bar components appears too
distinct between the two galaxy types.  This holds irrespective of the environment, since the frequency of bar components fitted in late types
does not change from low to high local galaxy density (see, however, \citealt{2012ApJ...761L...6M}). However, the shift of the decomposition parameters of late-type (cf. also \citealt{gut}) 
 towards those  of early-type dwarfs across the environments
suggests that environmentally induced processes could sufficiently alter the decomposition
parameters.

In the harassment simulations of \citet{2009A&A...494..891A}  massive disc galaxies
with bulge and disc components were transformed by several high-speed encounters with
other galaxies. During this process the bulge components  approximately retained their mass, 
but grew in size. Meanwhile the discs were truncated (see also \citealt{Mastropietro:2005ba}).
In the harassment simulations of \citet{Mastropietro:2005ba} bars were tidally induced, and
the final remnant was more spherical  (see also \citealt{lokas14,lokas16}). A more detailed suite of simulations is needed
to make firm predictions for the net effect on the frequency of bars.
While the disc truncation would make the outer component effectively smaller, 
shrinking  the inner component  by similar amounts seems much harder in this context, although in
some cases mass transfer via bars could work in this direction.
However, effective harassment comes with substantial mass loss. While this might be an option
for more massive disc galaxies (although probably on longer time-scales, see \citealt{Smith15};
and also  \citealt{2016arXiv160200527A} on the multicomponent aspect) this is unlikely for
late-type galaxies, which have similar stellar masses as the early-type dwarfs to start with.

Ram pressure stripping  \citep[e.g.][]{Boselli14} on the other hand is a more gradual process.
It can possibly lead to smaller sizes of both components, since it proceeds outside in. 
Some simulations even suggest that during the process central star formation
can be enhanced \citep[e.g.][]{Roediger}.
While it remains to be seen, whether the effect is large enough, the scenario is consistent with some of the early-type
dwarfs in the Virgo cluster having residual star formation in their centres \citep{liskerb}.
Another question is whether ram pressure stripping can explain the lack of elongated structures in early-type dwarfs.
For that  to work, these structures would probably need to be rather artefacts of somewhat irregular star 
formation, which can disappear after the cessation thereof, than dynamically stable structures.
A  possible imprint  of ram pressure stripping could be gradual changes in the stellar populations, which may be 
manifested in characteristic changes of the decomposition parameters
as a function of wavelength.

\section{Summary}
In order to test the hypothesis that early-type dwarfs are
transformed from  late-type spirals, we have analysed the
structural parameters of 500 late-type  galaxies (Hubble type $T\ge6$) in
the stellar mass range of about $10^7$ to a few times $10^{10}$ M$_\odot$ from the Spitzer Survey of
Stellar Structure in Galaxies (S$^4$G; \citealt{2010PASP..122.1397S}).  For the
structural parameters we use the two-dimensional multicomponent decompositions
from \citet{arXiv:1503.06550}. These decompositions are compared with those
previously made in a similar manner, and also in the near-infrared, for  early-type dwarfs in the
Virgo cluster by \citet[][]{2013arXiv1308.6496J}.

We found an increase of the fraction of galaxies that were well-fitted  with a 
single component with decreasing luminosity.
This trend is very similar for both early and late-type galaxies.
However, it might rather be driven by factors related to galaxy mass than being
a definitive sign that early-type dwarfs are transformed late-type galaxies.
Moreover, the structural components in late-type systems are on average
twice as large as those in early-type dwarfs, ruling out a simple fading.
 We discussed how  harassment and ram pressure can
potentially alter the structural decompositions, and 
 identified probable observable imprints. However, more detailed simulations
 need to test whether  transformations can yield the observed early-type
  decompositions parameters and which process contributes at what level.
  Nonetheless, the  changes of  late-type decomposition parameters from low to high galaxy density environments 
    suggest that such environmentally driven processes can contribute to the population of early-type dwarfs.

\section*{Acknowledgements}
We  thank the referee for a thorough report and suggestions,
which helped to improve the Letter. 
JJ  thanks the ARC for financial support via DP130100388. 
 EL, HS,  JL  acknowledge support from Academy of Finland, and
financial support from the DAGAL network (Marie Curie Actions), 
JL in addition from the Vilho, Yrj\"o ja Kalle V\'ais\"al\"a foundation of
the Finnish Academy of Science and Letters. 
TL was supported within the framework of the Excellence Initiative by 
the German Research Foundation (DFG) through the Heidelberg Graduate 
School of Fundamental Physics (grant number GSC 129/1).
This publication makes use of data products from the Two Micron All Sky Survey, which is a joint project of the University of Massachusetts and the Infrared Processing and Analysis Center/California Institute of Technology, funded by the National Aeronautics and Space Administration and the National Science Foundation.

\bibliographystyle{mn}

\label{lastpage}

\end{document}